\documentclass[iop]{emulateapj}
\shorttitle{A neutron star in J1018}
\shortauthors{Strader \etal~}
\def\etal{{et al.}}

\def\arcsec{\char'175 }

\def\hub{\ifmmode H_\circ\else H$_\circ$\fi}
\usepackage{subfigure}
\usepackage{amsmath}
\usepackage{graphics}
\usepackage{hyperref}
\usepackage{breakurl} 

\def\ltsima{$\; \buildrel < \over \sim \;$}
\def\simlt{\lower.5ex\hbox{\ltsima}} 
\def\gtsima{$\; \buildrel > \over \sim \;$}
\def\simgt{\lower.5ex\hbox{\gtsima}} 
\def\arcsec{\hbox{$^{\prime\prime}$}}

\begin{document}

\title{Optical spectroscopy of the high-mass $\gamma$-ray binary 1FGL J1018.6--5856: A probable neutron star primary}
\author{
Jay Strader\altaffilmark{1},
Laura Chomiuk\altaffilmark{1},
C.~C.~Cheung\altaffilmark{2},
Ricardo Salinas\altaffilmark{1,3},
Mark Peacock\altaffilmark{1}}

\altaffiltext{1}{Department of Physics and Astronomy, Michigan State University, East Lansing, MI 48824, USA}
\altaffiltext{2}{Space Science Division, Naval Research Laboratory, Washington, DC 20375, USA}
\altaffiltext{3}{Gemini Observatory, Casilla 603, La Serena, Chile}

\begin{abstract}

We present medium-resolution optical spectroscopy with the SOAR telescope of the O star secondary of the high-mass $\gamma$-ray binary 1FGL J1018.6--5856 to help determine whether the primary is a neutron star or black hole. We find that the secondary has a low radial velocity semi-amplitude of 11--12 km s$^{-1}$, with consistent values obtained for H and He absorption lines. This low value strongly favors a neutron star primary: while a black hole cannot be excluded if the system is close to face on, such inclinations are disallowed by the observed rotation of the secondary. We also find the high-energy (X-ray and $\gamma$-ray) flux maxima occur when the star is behind the compact object along our line of sight, inconsistent with a simple model of anisotropic inverse Compton scattering for the $\gamma$-ray photons.

\end{abstract}
 
\keywords{pulsars: general --- Gamma rays: general --- X-rays: general --- binaries: spectroscopic --- stars: individual (1FGL J1018.6--5856)}

\section{Introduction\label{section-intro}}

There are five known ``$\gamma$-ray binaries": high-mass X-ray binaries with variable $\gamma$-ray emission and very high energy (VHE, defined at $> 100$ GeV) detections (Dubus 2013). The most recent detection, 1FGL J1018.6--5856, was made by \emph{Fermi}-LAT and consists of a massive O star secondary in orbit with a neutron star or black hole at a distance of $\sim 5$ kpc (Corbet \etal~2011; Ackermann \etal~2012; Napoli \etal~2011). The $\gamma$-ray and X-ray flux are modulated on a period of about 16.5 d, which is interpreted as the orbital period (Ackermann \etal~2012; An \etal~2015).

The two basic models for $\gamma$-ray emission from $\gamma$-ray binaries alternatively invoke a neutron star or black hole as the compact object. In the former case the $\gamma$-ray emission likely originates in the interaction between the pulsar wind and the stellar wind (and/or disk) of the companion. In the latter case the black hole is assumed to be accreting and the high-energy emission is associated with a jet. In only one case (the neutron star PSR B1259--63) is the nature of the compact object definitively known. These scenarios are reviewed in detail by Dubus (2013), who argues that the balance of the evidence favors the pulsar model for known $\gamma$-ray binaries.

For 1FGL J1018.6--5856, there is only indirect evidence for the identification of its primary star. Waisberg \& Romani (2015) used optical spectroscopy to suggest the O star had an orbital semi-amplitude in the range 15--40 km s$^{-1}$, with the higher end favoring a black hole. Williams \etal~(2015) discussed binary evolution simulations of the formation of 1FGL J1018.6--5856, which favor a heavy neutron star as the primary. These same simulations also predict a large eccentricity that has not (yet) been observed.

Here we present new optical spectroscopy of 1FGL J1018.6--5856 that allows a clean measurement of the orbital semi-amplitude and good constraints on the nature of the compact object in the binary.

\section{Observations\label{sec-observations}}

All spectroscopic observations were obtained using the Goodman High-Throughput Spectrograph (Clemens \etal~2004) on the SOAR 4.1-m telescope, comprising 14 epochs from UT 2014 Dec 12 to 2015 Aug 25. We used a 1.03\arcsec\ slit and a 2400 l mm$^{-1}$ grating (resolution 0.8 \AA), with an approximate wavelength range 4260--4830 \AA. We reduced the spectra in the usual manner, with optimal extraction and wavelength calibration using FeAr arcs taken after each set of two to three 5--10 min exposures.

\section{Results and Discussion}

\subsection{Radial Velocities}

The strongest line apparent in the 1FGL J1018.6--5856 spectra is H$\gamma$, with \ion{He}{2} lines at 4542 \AA\ and 4686 \AA\ also clear. The \ion{He}{1} line at 4471 \AA\ is also seen, but is weaker than the other lines mentioned. There is \ion{N}{3} emission visible in the higher signal-to-noise spectra. Absorption observed at 4430 \AA\ and 4762 \AA\ is due to interstellar dust (e.g., Sota \etal~2011). The spectrum looks very similar to that plotted in Ackermann \etal~(2012).

To measure barycentric radial velocities, we cross-correlated the object spectra with a spectrum of the O6V star HD 172275 (Wegner 2002) taken with the same setup.  	
Waisberg \& Romani (2015) found different results for the radial velocity of 1FGL J1018.6--5856 between H and He lines, so we also consider these lines separately: first we performed the cross-correlation in the region around H$\gamma$, then did the same for the \ion{He}{2} lines. Given the long period of the system, at each epoch we take a weighted average of the velocities derived from the 2--3 individual spectra obtained over 15--30 min, so there are 14 radial velocity measurements total for each of H$\gamma$ and \ion{He}{2}. These are the values listed in Table 1.

We find weak evidence for a systematic difference in the radial velocities derived from these sets of lines: the H velocities are in the median $6\pm2$ km s$^{-1}$ smaller than those from \ion{He}{2}. This difference is in the same direction, but of a much smaller magnitude, than the $\sim 20$ km s$^{-1}$ offset measured between H and \ion{He}{2} lines in the O6.5V((f)) secondary in the $\gamma$-ray binary LS 5039 (Sarty \etal~2011). This difference is thought to be due to wind opacity. For the purpose of this paper, we simply note that for 1FGL J1018.6--5856 the measured velocity differences are not correlated with the velocities themselves, so the offset only affects the derived systemic velocity, not the semi-amplitude, and hence does not affect conclusions about the nature of the compact object in the system. The uncertainties in the \ion{He}{2} measurements are larger, so for the remainder of the paper (with two exceptions below) we use the H velocities for our analysis, but list both sets of values in Table 1.

\begin{deluxetable}{lcccc}
\tablecaption{Radial Velocities \label{tab:dat1}}
\tablehead{BJD\tablenotemark{a}    & Vel. (H$\gamma$) & Vel. (He II) & $\phi$\tablenotemark{b}  (H$\gamma$) & $V_{\rm r} \, {\rm sin} \, i$\tablenotemark{c} \\
                  (days) & (km s$^{-1}$)        & (km s$^{-1}$)   &  & (km s$^{-1}$) }
\startdata
2457003.7584176 & $30.9\pm4.6$ & $48.2\pm8.8$ & 0.427 & 195  \\
2457012.8342719 & $21.5\pm4.3$ & $33.0\pm7.7$ & 0.975 & 207 \\
2457022.7931137 & $38.5\pm3.7$ & $47.5\pm4.8$ & 0.577  & 270 \\
2457037.8082448 & $40.2\pm3.9$ & $46.3\pm6.5$ & 0.485  & 206 \\
2457071.7659660 & $46.1\pm4.1$ & $61.8\pm7.1$ &  0.538  & 205 \\
2457120.6160817 & $43.4\pm3.9$ & $45.2\pm7.3$ &  0.490  & 276 \\
2457158.5171730 & $24.2\pm4.1$ & $17.2\pm7.2$ &  0.781  & 266 \\
2457166.5471461 & $31.7\pm4.0$ & $44.8\pm7.5$ &  0.267  & 289 \\
2457170.5363727 & $40.1\pm3.8$ & $41.1\pm6.9$ &  0.508  & 261 \\
2457186.5474504 & $38.0\pm3.7$ & $44.1\pm7.5$ &  0.475  & 266 \\
2457195.5188204 & $16.8\pm4.5$ & $23.3\pm7.5$ &  0.018  & 243 \\
2457252.4924565 & $55.4\pm4.5$ & $41.1\pm6.9$ &  0.461  & 261 \\
2457257.4695183 & $32.1\pm5.0$ & $28.8\pm7.6$ &  0.762  & 274 \\
2457260.4750679 & $22.8\pm5.9$ & $26.4\pm9.4$ &  0.944  & 277 \\
\enddata
\tablenotetext{a}{Barycentric Julian Date on the TDB system of the midpoint of the velocities.}
\tablenotetext{b}{Phase of observation, defined with respect to the ascending node of the compact object at $\phi = 0$. 
Those for \ion{He}{2} are formally larger by 0.084 due to a different inferred time of ascending node (\S 3.2).}
\tablenotetext{c}{Inferred projected rotational velocity, as discussed in \S 3.3.}
\end{deluxetable}

\subsection{Spectroscopic Orbit}

We fit a standard Keplerian model to the H$\gamma$ radial velocities in Table 1 after correcting the observation midpoints to Barycentric Julian Date (BJD) on the Barycentric Dynamical Time (TDB) system (Eastman \etal~2010). The period was fixed to the best-fit X-ray period of $P=16.544$ d (An \etal~2015), though we note that even with our modest phase coverage, if the period is left free, the best-fit value was $16.4$ d, consistent with the high-energy period. Initially we fixed the eccentricity ($e$) to zero. We found a reasonable fit ($\chi^2$ = 19/11 d.o.f.; rms 5.2 km s$^{-1}$) for semi-amplitude $K_2 = 11.4\pm1.5$ km s$^{-1}$, systemic velocity $v_{sys} = 30.4\pm1.3$ km s$^{-1}$ and the BJD time of the ascending node of the compact object ($\phi =0$): $T_{0}$ = $2457244.86\pm0.49$. This fit is shown in Figure 1 (top). The true uncertainty for the systematic velocity is larger than listed (realistically, perhaps 10 km s$^{-1}$) considering the uncertainty in the barycentric velocity of HD 172275 and the differences found between the H and He lines as discussed above. For the \ion{He}{2} lines, $v_{sys} = 36.2\pm2.2$ km s$^{-1}$.

Leaving the eccentricity free does not significantly improve the fit; on the other hand, fixing the eccentricity at a range of higher values ($e \sim 0.1$--0.5) gives fits nearly indistinguishable in quality. Thus the eccentricity is essentially unconstrained with these data: even very large values are not definitively excluded.

We also performed circular fits instead using the \ion{He}{2} velocities. The best-fit model is plotted in Figure 1 (bottom). As expected given the systematic offset from H$\gamma$, the derived systemic velocity was significantly different, but the best-fit semi-amplitude was $K_2 = 12.2\pm2.7$ km s $^{-1}$, in excellent agreement with the value above. The much higher value of $K_2$ found by Waisberg \& Romani (2015) from the \ion{He}{2} lines in their spectra (up to 40 km s$^{-1}$) thus appears to have been spurious. For the reasons stated above we use the H$\gamma$ radial velocities for the analysis below, but there would be no substantial change to our conclusions about the identity of the compact object if the \ion{He}{2} velocities were used instead.

The time of the ascending node of the compact object, and hence that of both conjunctions, is smaller by 1.39 d for the \ion{He}{2} fit compared to the H$\gamma$ fit. This corresponds to a phase difference of $0.084$ and hence is relevant for the interpretation of the high-energy observations (\S 3.5). The formal random uncertainty for each measurement is about 0.5 d, so an offset of 1.4 d is larger than would be expected. As discussed above, the differences between the H$\gamma$ and \ion{He}{2} velocities are likely to be dominated by systematic effects, so we do not average the respective $T_{0}$ measurements. However, we caution that the uncertainty in the relative phase of conjunction may be somewhat larger than that expected from the uncertainty in $T_{0}$ alone ($\sim 0.03$ in phase).
 
The reader may note that there is a clear outlying data point in Figure 1 for H$\gamma$ (at $\sim 55$ km s$^{-1}$). If this point is excluded, the quality of the fit is substantially improved and the inferred semi-amplitude is about 0.8 km s$^{-1}$ lower. There is nothing else abnormal about the spectra at this epoch, so in the absence of additional information we retain all data. Nonetheless the subsequent analysis may be taken as conservative, in that the evidence for a neutron star may be marginally stronger than presented below.

\begin{figure}[ht]
\includegraphics[width=3.3in]{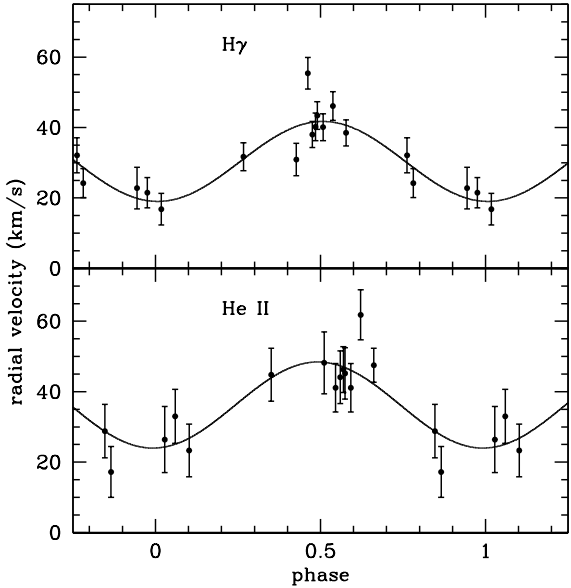}
\caption{Orbital fits to radial velocities of the secondary of 1FGL J1018.6--5856, showing H$\gamma$ (top) and \ion{He}{2} (bottom). In our phase convention the O star is behind the compact object at $\phi = 0.75$.}
\end{figure}

\subsection{Rotational Velocity and Inclination Limits}

While not designed for this purpose, our observations provide some constraints on the projected rotational velocity ($V_{\rm r} \, {\rm sin} \, i$) of the O star. We convolved the spectrum of the comparison star HD 172275 with a set of kernels reflecting a range of rotational velocities, assuming a standard limb darkening law with coefficient $\epsilon = 0.23$. We then cross-correlated these spectra with the original spectrum in the region around the two \ion{He}{2} lines as a compromise choice between lines likely to primarily reflect photospheric motions and those with sufficient signal-to-noise for this measurement. This produced a relation between the measured full-width of half maximum (FWHM) and $V_{\rm r} \, {\rm sin} \, i$. We then cross-correlated the spectra of 1FGL J1018.6--5856 at each epoch with HD 172275 in the same wavelength regions and converted the resulting FWHM measurements into estimates of $V_{r} \, {\rm sin} \, i$. These are listed in Table 1. All of these measurements should be lower limits as they assume the (unmeasured) $V_{\rm r} \, {\rm sin} \, i$ of HD 172275 itself is negligible. 

The  $V_{\rm r} \, {\rm sin} \, i$ estimates are bimodal: at some epochs the value is $\sim 205$ km s$^{-1}$ and at some epochs higher at $\sim 270$ km s$^{-1}$. The differences do not correlate with the phase of observation. We assume the lower value better reflects the true projected rotational velocity but recognize that these lines do not trace the photosphere with fidelity, possibly due to the effects of a time-variable wind (Sarty \etal~2011; Waisberg \& Romani 2015). Perhaps consistent with this hypothesis, the lower $V_{\rm r} \, {\rm sin} \, i$ estimates all occurred in the time range 2014 Dec to 2015 Feb (though one
larger value was also inferred within this range), while all measurements from 2015 Apr to 2015 Aug were larger.

The measurement of rotation sets a \emph{lower} limit on the inclination of the system due to the maximum (breakup) velocity of O stars of various masses. Using the stellar parameters in Martins \etal~(2005), the critical velocity varies from $\sim 710$ to 765 km s$^{-1}$ depending on mass. Assuming $V_{\rm r} \, {\rm sin} \, i >$ 205 km s$^{-1}$, then the inclination $i \gtrsim 15^{\circ}$, ruling out very face-on inclinations.

\subsection{Mass of the Compact Object}

The standard formula for the ``mass function" $f(M_1)$ of a circular single-lined spectroscopic binary is: $f(M_1) = P K_2^3/(2 \pi G) = (M_1 \, \textrm{sin} \, i)^3/(M_1+M_2)^2$, for inclination $i$ and secondary mass $M_2$. Using the values above, $f(M_1) = 0.0025\pm0.0010$. Waisberg \& Romani (2015) cite Casares \etal~(2005) for the mass range of an O6V((f)) star as 20.0--26.4 $M_{\odot}$ based on the mass inferred for the similar star in the $\gamma$-ray binary LS 5039. It is then straightforward to determine the mass $M_1$ of the compact object as a function of inclination.

For $M_2 = 20 M_{\odot}$ and a standard neutron star mass in the range 1.4--2.0 $M_{\odot}$, $i$ must be between $49^{\circ}$ and $32^{\circ}$, with the lower limit dropping to $26^{\circ}$ if an upper mass limit of 2.5 $M_{\odot}$ is used. For the larger value of $M_2 = 26.4 M_{\odot}$, the constraints are naturally weaker, with the neutron star mass range 1.4--2.0 $M_{\odot}$ corresponding to $i$ from $64^{\circ}$ and $39^{\circ}$. If the binary evolution modeling of Williams \etal~(2015) is accurate, favoring a heavy neutron star, then the inclination is most likely to be in the range $\sim 25$--40$^{\circ}$.

Stellar-mass black holes have typical masses $\gtrsim 5 M_{\odot}$ (e.g., Farr \etal~2011). This primary mass would be allowed by the spectroscopic observations only if $i \lesssim 13^{\circ}$--$16^{\circ}$ ($M_2 = 20$--26.4 $M_{\odot}$). Due to the presence of (sin $i$)$^3$ in the mass function equation, this conclusion would be very similar even if the secondary were somewhat more massive ($\sim 30 M_{\odot}$; Napoli \etal~2011). However, these same inclinations are generally excluded by the measurement of the projected rotational velocity of the star (\S 3.3). 

The straightforward conclusion from these data is that the low semi-amplitude of the secondary provides good evidence in favor of identifying the primary as a neutron star. Adding in the (less secure) 
$V_{r} \, {\rm sin} \, i$ measurement eliminates nearly all combinations of secondary mass and inclination that would allow a black hole. We note that the constraints on the compact object published in Waisberg \& Romani (2015) are not correct in detail owing to a plotting error in their Figure 5 (R.~Romani, private communication).

Future spectroscopic observations are unlikely to substantially change the measured value of the semi-amplitude, but could allow an improved lower limit on the inclination and definitively rule out a black hole. Better spectroscopic phase coverage would allow improved constraints on the eccentricity, which would help in modeling the origin of the system and its $\gamma$-ray emission.

\subsection{Phase of high energy emission}

If the orbitally-modulated $\gamma$-ray emission is due to anisotropic inverse Compton scattering and the system has modest eccentricity, the $\gamma$-ray flux is expected to peak when the star is in front of the compact object  along our line of sight, with a stronger effect for more edge-on inclinations (Ackermann \etal~2012). At higher $\gamma$-ray energies (approaching the VHE regime) pair production at similar orbital phases can reduce the flux and soften the spectrum. On the basis of the VHE spectrum and the similarity of the \emph{Fermi} and H.E.S.S. light curves, Abramowski \etal~(2015) argue that 1FGL J1018.6--5856 must be a relatively low-inclination, low-eccentricity system.

Ackermann \etal~(2012) found that the peak of the GeV $\gamma$-ray emission in the phase-binned light curve (and the time when the spectrum is the hardest) occurs at a phase $\phi=0.72$ in our convention (where $\phi=0.75$ is when the star is behind the compact object\footnote{In Ackermann \etal~(2012) the $\gamma$-ray maximum is denoted as $\phi=0$ since the geometry of the binary was unknown.}). This phase is uncertain by at least 0.04 due to the combined uncertainties in the $\gamma$-ray maximum and the time of conjunction (with the latter possibly even more uncertain; \S 3.2). The X-ray and VHE emission also peak at a similar phase (An \etal~2015; Abramowski \etal~2015). Even given the phase uncertainties, the data are consistent with the X-ray, GeV and VHE maxima occurring when the star is behind the compact object. This inference is unexpected compared to the simple model of anisotropic inverse Compton scattering discussed above. A possible solution---though one perhaps at odds with the VHE light curve---would be if the orbit were in fact eccentric and periastron occurred at a similar phase. In any case, this is a motivation for improving the spectroscopic measurement of the both the orbital eccentricity and the phase of conjunction.

\acknowledgments

We thank an anonymous referee for helpful comments that improved the paper. Based on observations obtained at the Southern Astrophysical Research (SOAR) telescope, which is a joint project of the Minist\'{e}rio da Ci\^{e}ncia, Tecnologia, e Inova\c{c}\~{a}o (MCTI) da Rep\'{u}blica Federativa do Brasil, the U.S. National Optical Astronomy Observatory (NOAO), the University of North Carolina at Chapel Hill (UNC), and Michigan State University (MSU). Support from NASA grant NNX15AU83G is gratefully acknowledged. Work by C.C.C. at NRL is supported in part by NASA DPR S-15633-Y.

{}

\end{document}